\documentclass[aps,12pt, a4paper]{revtex4}

\usepackage[utf8]{inputenc}
\usepackage[T1]{fontenc}
\usepackage{amsmath}
\usepackage{amsfonts}
\usepackage{amssymb}
\usepackage{graphics,graphicx}

\usepackage{graphicx}
\usepackage{subfigure} 
\usepackage{wrapfig} 
\usepackage{epstopdf}
\usepackage{xcolor}
\graphicspath{{figuras/}}

\usepackage[breaklinks=true]{hyperref}
\usepackage{setspace}

\begin{document}

\title{Modeling dark sector in Horndeski gravity at first-order formalism}
\author{F. F. Santos $^{1}$, R.M.P. Neves$^{1}$ and F. A. Brito$^{1,2}$}
\email{fabiano.ffs23@gmail.com, raissaeter@yahoo.com.br, fabrito@df.ufcg.edu.br}

\affiliation{$^1${Departamento de F\'\i sica, Universidade Federal da Para\'iba, Caixa Postal 5008, 58051-970 Jo\~ao Pessoa PB, Brazil} }
\affiliation{$^2${Departamento de F\'\i sica, Universidade Federal de Campina
Grande, Caixa Postal 10071, 58109-970, Campina Grande PB, Brazil} }
\date{\today}

\begin{abstract}
We investigate a cosmological scenario by finding solutions using first-order formalism in the Horndeski gravity that constrains the superpotential and implies that no free choice of scalar potential is allowed. Despite this we show that a de Sitter phase at late-time cosmology can be realized, where the dark energy sector can be identified.  The scalar field equation of state tends to the cosmological scenario at present time and allows us to conclude that it can simulate the dark energy in the Horndeski gravity.
\end{abstract}

\maketitle
\newpage

\tableofcontents
\newpage

\section{Introduction}

Since Einstein proposes the General Relativity it has been supported by strong observation evidence in many astrophysical scenarios, namely, the Eddington's measurement of the deflection of light in 1919 and recent direct observation of gravitational waves by the LIGO collaboration \cite{Papallo:2019lrl,Abbott:2016blz}. However, we still have fundamental problems to be well-understood in General Relativity such as dark matter, dark energy and inflationary phase of the Universe. In recent years one has been proposed models involving modifications of General Relativity \cite{Zumalacarregui:2013pma,Cisterna:2017jmv}. In such modifications are maintained some of its essential properties such as a second order of the equations of motion arising from a diffeomorphism-invariant action and keeping the Lorentz invariance. Due to these assumptions, the additional propagating degrees of freedom into the gravity sector consists of including additional fields (scalars, vectors or tensors) \cite{Heisenberg:2018vsk}. These modifications of the gravity theory taking into account nonminimal couplings between geometry and matter become one of the mainstream of modified gravity theories and the applications of the nonminimal couplings of matter with gravity provides a way to solve the cosmological constant problem \cite{Dolgov:2003fw,Mukohyama:2003nw} and accelerated expansion of the Universe \cite{Nojiri:2004bi,Allemandi:2005qs} --- see also alternative theories, for instance, involving late-time \cite{brito2005,brito2007} and early-time acceleration (inflation) \cite{Santos} in the context of supergravity.

Recent investigations about Einstein gravity have called attention for the coupling of the theory to scalar fields \cite{Brito:2019ose}. These efforts led to the development of the well-known Galileons that are scalar-tensor theories \cite{Nicolis:2008in}. Indeed, these studies have led to the rediscovery of the Horndeski gravity.

The Horndeski gravity was originally discovered in 1974  \cite{Horndeski:1974wa,Anabalon:2013oea,Cisterna:2014nua,Brito:2018pwe,Heisenberg:2018vsk}. It is a general single scalar field-tensor theory with second-order field equations and second order energy-momentum tensor. The Lagrangian producing second order equations of motion as discussed in  \cite{Deffayet:2011gz,Anabalon:2013oea,VanAcoleyen:2011mj,Gomes:2015dhl,Rinaldi:2016oqp,Cisterna:2017jmv,Cisterna:2014nua} includes four arbitrary functions of the scalar field and its kinetic term \cite{Gleyzes:2013ooa,Zumalacarregui:2013pma}. The term that we are interested in includes a nonminimal coupling between the standard scalar kinetic term and the Einstein tensor. Besides the cosmological interest, recent investigation has also been called attention in astrophysics, such as the searching for black hole solutions which develop Hawking-Page phase transitions at a critical temperature \cite{Anabalon:2013oea}. Other examples of spherically symmetric solutions in Horndeski theory in the context of the solar system and further astrophysical scenarios can also be found, for instance, in the study of perihelion shift and light bending \cite{Bhattacharya:2016naa} and in the issues involving properties of spinning gyroscope and the Gravity Probe B experiment \cite{Mukherjee:2017fqz}. {Some applications in astrophysical compact objects have also been considered in \cite{erices,Cisterna:2016vdx}.}

An interesting problem in the cosmological scenario is the cosmological constant problem that is related to the discrepancy between the natural scale and its measured value. As discussed in \cite{Babichev:2015qma}, it is possible to address this problem by using a self-tunning mechanism, which has been analyzed in the original Horndeski theory for the so-called Fab Four theory \cite{Charmousis:2011ea,Charmousis:2011bf}. However, more analysis of the cosmological self-tunning and the local solutions in the context beyond Horndeski theories can be found in \cite{Babichev:2016kdt}.

In the cosmological scenario, the Horndeski cosmological models are able to screen the vacuum energy coming from any field theory in a space that should be a de Sitter vacuum \cite{Martin-Moruno:2015bda,Martin-Moruno:2015eqa}. In these models, we can understand that the current accelerated expansion of the Universe is a dynamical result evolution of a de Sitter attractor \cite{Martin-Moruno:2015lha}. In this sense, the Horndeski models involving a de Sitter critical point for any kind of material content may provide a mechanism to alleviate the cosmological problem \cite{Martin-Moruno:2015kaa}. Thus, the models involving nonminimal derivative couplings to gravity has been explored in a variety of extended theories of gravity \cite{Deffayet:2010qz}. These models show peculiar features, for example, an essential mixing of scalar and tensor kinetic terms, named kinetic braiding, and possess a rich cosmological phenomenology that includes a late-time asymptotic de Sitter state that allows a phantom-divide line crossing with neither ghosts nor gradient instabilities.

In this work, we consider the Horndeski theory in the cosmological context by using the first-order equation formalism that was presented recently for Horndeski gravity in a braneworlds scenario \cite{Brito:2018pwe}. We investigate the second-order equations through the first-order formalism because in general one simplifies the study of analytical or numerical solutions. {Moreover, the first-order equation formalism is a fundamental tool in renormalization group (RG) flow in holographic cosmology \cite{McFadden:2009fg,Baumann:2019ghk,Kiritsis:2019wyk,Kiritsis:2013gia}}. In our case, we consider the Friedmann equations without curvature and assume dark energy dominance. In particular, the inflationary context was analyzed by considering a power-law potential and using the dynamical system method to investigate the possible asymptotical regimes of the model \cite{Skugoreva:2013ooa}. It was shown for sloping potentials there exists a quasi-de-Sitter phase corresponding to the early inflationary Universe. In our investigations, by considering numerical methods, we show that kink type solutions of first-order equations represent a de Sitter Universe. We address several important issues in cosmological observables, such as the Hubble function, the deceleration parameter, and the dark energy equation-of-state. We investigate their evolution at small redshifts for a general scalar potential written in terms of a superpotential.
Furthermore, as shown in  \cite{Casalino:2018mna} the Horndeski action in the Friedmann frame {\it without scalar potential} cannot describe the dark matter and dark energy, due the instability and also by constraint of gravity waves, the scalar field on the background evolution is negligible, and the presence of this field becomes unnecessary for explaining the dark matter and dark energy. In our case, however, the nature of the scalar potential in the Horndeski gravity in the Friedmann frame is much satisfactory for describing dark energy.

The paper is organized as follows.  In Sec.~\ref{z1}, we present the first-order formalism in Horndeski theory with a scalar potential given in terms of an implicit superpotential obtained numerically. In Sec.~\ref{z2}, we use the numerical method to find cosmological solutions that represent a de Sitter Universe. In Sec.~\ref{z3}, we discuss the following cosmological observables: the Hubble function, the deceleration parameter, and the dark energy equation-of-state at small redshifts. Finally, in Sec.~\ref{z5}, we present our conclusions. 

\section{The Horndeski gravity with a scalar potential}\label{z1}

In our present investigation, we shall address the study of Friedmann-Robertson-Walker (FRW) solutions in the framework of the Horndeski gravity \cite{Horndeski:1974wa,Cisterna:2014nua,Feng:2015oea,Anabalon:2013oea,Brito:2018pwe,Brito:2019ose} which action with a scalar potential reads 
\begin{equation}
I[g_{\mu\nu},\phi]=\int{\sqrt{-g}d^{4}x\left[kR-\frac{1}{2}(\alpha g_{\mu\nu}-\eta G_{\mu\nu})\nabla^{\mu}\phi\nabla^{\nu}\phi-V(\phi)\right]}.\label{10}
\end{equation}
Note that we have a nonminimal scalar-tensor coupling where we can define a new field $\dot{\phi}\equiv\psi$. This field has dimension of $(mass)^{2}$ and the parameters $\alpha$ and $\eta$ control the strength of the kinetic couplings, $\alpha$ is dimensionless and $\eta$ has dimension of $(mass)^{-2}$. Thus, the Einstein-Horndeski field equations can be formally written as in the usual way
\begin{equation}
G_{\mu\nu}=\frac{1}{2k}T_{\mu\nu}\label{11},
\end{equation}
where $T_{\mu\nu}=\alpha T^{(1)}_{\mu\nu}-g_{\mu\nu}V(\phi)+\eta T^{(2)}_{\mu\nu}$ with $k=(16\pi G)^{-1}$ and the scalar field equation is given by 
\begin{equation}
\nabla_{\mu}[(\alpha g^{\mu\nu}-\eta G^{\mu\nu})\nabla_{\nu}\phi]=V_{\phi}.\label{12}
\end{equation}
{We shall adopt the notation $f_{\phi\phi...\phi}(\phi)\equiv d^n f(\phi)/d\phi^n$. In particular, $V_\phi\equiv{dV}/{d\phi}$.}
The aforementioned energy-momentum tensors $T^{(1)}_{\mu\nu}$ and $T^{(2)}_{\mu\nu}$ take the following form
\begin{equation}\begin{array}{rclrcl}
T^{(1)}_{\mu\nu}&=&\nabla_{\mu}\phi\nabla_{\nu}\phi-\frac{1}{2}g_{\mu\nu}\nabla_{\lambda}\phi\nabla^{\lambda}\phi,\\
T^{(2)}_{\mu\nu}&=&\frac{1}{2}\nabla_{\mu}\phi\nabla_{\nu}\phi R-2\nabla_{\lambda}\phi\nabla_{(\mu}\phi R^{\lambda}_{\nu)}-\nabla^{\lambda}\phi\nabla^{\rho}\phi R_{\mu\lambda\nu\rho}\\
              &-&(\nabla_{\mu}\nabla^{\lambda}\phi)(\nabla_{\nu}\nabla_{\lambda}\phi)+(\nabla_{\mu}\nabla_{\nu}\phi)\Box\phi+\frac{1}{2}G_{\mu\nu}(\nabla\phi)^{2}\\
							&-& g_{\mu\nu}\left[-\frac{1}{2}(\nabla^{\lambda}\nabla^{\rho}\phi)(\nabla_{\lambda}\nabla_{\rho}\phi)+\frac{1}{2}(\Box\phi)^{2}-(\nabla_{\lambda}\phi\nabla_{\rho}\phi)R^{\lambda\rho}\right].\label{g}
\end{array}\end{equation}
Here we are interested in investigating the cosmological implications of theories with extended nonminimal derivative couplings. Considering the flat FRW metric of the form
\begin{equation}
ds^{2}=-dt^{2}+a^{2}(t)\delta_{ij}dx^{i}dx^{j},\label{13}
\end{equation}
the scalar field depends on the cosmic time only and computing the $tt$-component of the Einstein-Horndeski field equations (\ref{11}) gives
\begin{equation}
3\dot{a}(t)^{2}(4k-3\eta\psi^{2}(t))-a^{2}(t)[\alpha\psi^{2}(t)+2V(\phi)]=0.\label{2i}
\end{equation}
The Friedmann equation can be readily found from this equation and reads
\begin{equation}
H^{2}=\frac{\alpha\dot{\phi}^{2}+2V(\phi)}{3(4k-3\eta\dot{\phi}^{2})}.\label{jj}
\end{equation}
Now we use the first-order formalism by assuming \cite{brito2007, Bazeia:2005tj, Bazeia:2006mh}
\begin{eqnarray}\label{1st-order-1}
H&=&W(\phi),\qquad H\equiv\frac{\dot{a}}{a}\nonumber\\
 \dot{\phi}&=&-W_{\phi}(\phi),
\end{eqnarray}
where the superpotential $W(\phi)$ plays a central role. Through these equations and equation (\ref{jj}), we can write the scalar potential   as follows
\begin{equation}
V(\phi)=\frac{3}{2}W^{2}(4k-3\eta W^{2}_{\phi})-\frac{\alpha}{2}W_{\phi}^{2}.\label{1x}
\end{equation}
Notice the scalar potential is similar to that found in the braneworlds scenario \cite{Brito:2018pwe}. Now we proceed with the $xx$-$yy$-$zz$-components of the equation (\ref{11}) which are given by
\begin{equation}
\dot{a}^{2}(t)(\eta\psi^{2}(t)-4k)+a(t)[\ddot{a}(t)(2\eta\psi^{2}(t)-8k)+4\eta\dot{a}(t)\psi(t)\dot{\psi}(t)]+a^{2}(t)[\alpha\psi^{2}(t)+2V(\phi)]=0.\label{3i}
\end{equation}
The scalar field equation (\ref{12}) in FRW background is written in the form
\begin{equation}
\ddot{\phi}+3H\dot{\phi}+\frac{6\eta\dot{\phi} H\dot{H}}{\alpha+3\eta H^{2}}+\frac{V_{\phi}(\phi)}{\alpha+3\eta H^{2}}=0,\label{2d}
\end{equation}
and for $V_{\phi}=0$ reduces to the form found in \cite{Rinaldi:2016oqp}. Now combining the equation (\ref{2i}) with equation (\ref{3i}) we can write a differential equation for the superpotential
\begin{eqnarray}
2WW_{\phi\phi}+W_{\phi}^{2}+3W^{2}-\beta=0,\label{R2} 
\end{eqnarray}
where $\beta=(1-\alpha)/\eta$. There is a family of analytical solutions for the homogeneous case $\beta=0$ given in terms of trigonometric functions 
\begin{eqnarray}\label{W-an}
W(\phi)=\left(-C_1\sin{\left(\frac{3}{2}\phi\right)}+C_2\cos{\left(\frac{3}{2}\phi\right)}\right)^{2/3}
\end{eqnarray}
From equations (\ref{1st-order-1}) we can find an approximated solution, for $C_1=0, C_2=1$ and, e.g., in the limit of one of the functions is very small. Here we assume $\phi\approx1$ for a cosmological time around a time scale $t^*$, to find an acceptable cosmological solution for accelerating Universe
\begin{eqnarray}\label{1st-order-phantom}
\phi&\approx& 2.41 t-\phi_0\nonumber\\
a(t)&\approx& \exp{(0.17t)}.
\end{eqnarray}
As we shall see below, for arbitrary values of $\beta\neq0$, we should apply numerical methods to find exact solutions for the equations (\ref{1st-order-1}) and  (\ref{R2}). 


\section{Numerical solutions}\label{z2}

\begin{figure}[!ht]
\begin{center}
\includegraphics[scale=0.35]{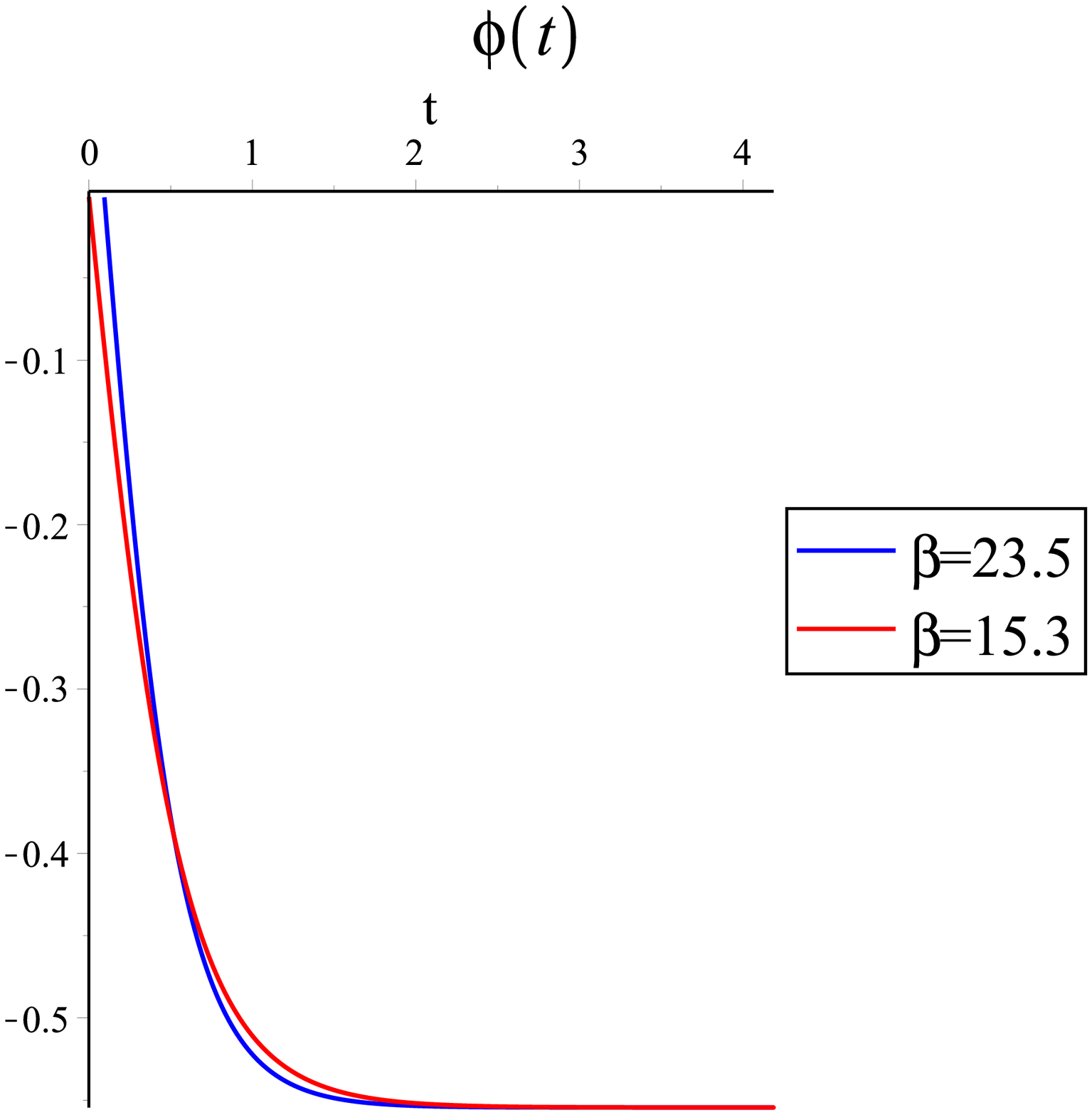}
\includegraphics[scale=0.6]{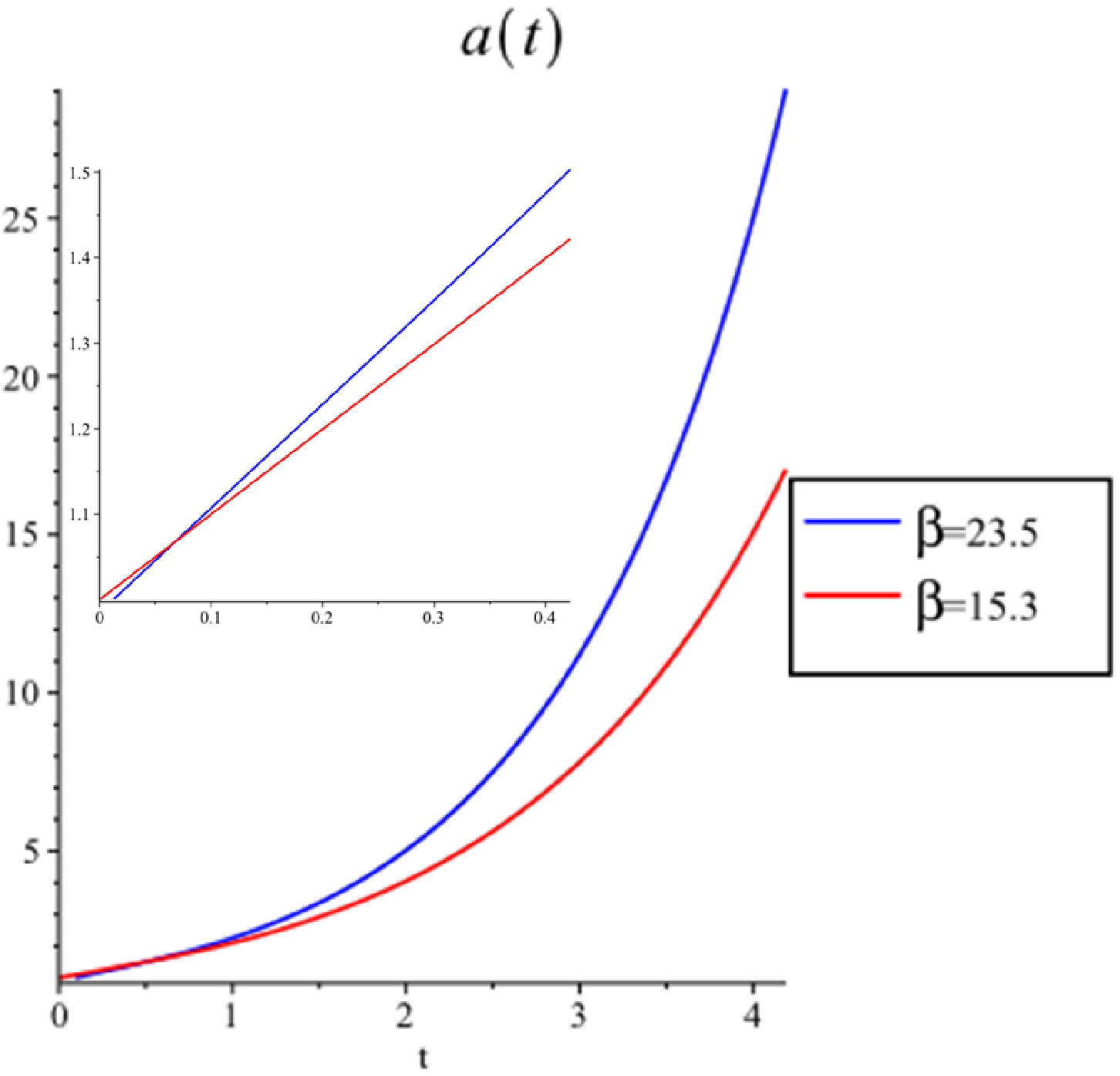}
\caption{The behavior of the solutions of equation (\ref{1st-order-1}): $\phi(t)$  (left panel) 
for $\beta=23.5$, with $\alpha=0.06$, $\eta=0.04$ (blue curve) and $\beta=15.3$, with $\alpha=0.08$, $\eta=0.06$ (red curve); and $a(t)$ (right panel) for the same values of parameters. The inset shows the behavior of $a(t)$ at smaller times. Particularly, $a(0)=1$, for $\beta=15.3$ (red curve), {where $t=0$ does not mean the Big-Bang. Instead, it means the time at which our formalism starts to describe
the present phase of the Universe}.}
\label{z02}
\end{center}
\end{figure}

The pair of first-order equations can be solved numerically for a broader range of Horndeski parameter values as long as we assume appropriate boundary conditions. In Fig.~\ref{z02} we show the behavior of the scalar field with a `kink' profile (left panel) and scale factor (right panel) associated with the FRW solutions for $\beta=23.5$ with $\alpha=0.06$, $\eta = 0.04$ (blue curve) and $\beta=15.3$ with $\alpha=0.08$, $\eta= 0.06$ (red curve). 
These regimes of small $\eta$ and $\alpha\neq1$ are required to produce acceptable cosmological solutions. Of course, these choices of parameters lead to a non-homogeneous limit of equation (\ref{R2}) where no analytical solutions are known. Thus, we shall find such cosmological solutions by using numerical methods. In the present case, we have applied the Runge-Kutta method to first-order equations (\ref{1st-order-1}) for $a(t)$ and $\phi(t)$ and second-order equation (\ref{R2}) for $W(\phi)$. 
The boundary conditions we used here were the following: $W(0)=1$, $W'(0)=1$, $\phi(0)=0$. 

 
\section{Cosmological Observables}\label{z3}

In this section we investigate cosmological observables in our cosmological setup in Horndeski gravity. Several studies in this context have already been considered and shown to produce successful models \cite{Gomes:2015dhl,Rinaldi:2016oqp,Harko:2016xip}, as for example in the description of dark energy \cite{Rinaldi:2016oqp}. 

 {We extend these earlier studies in the context of the first-order formalism, which is in the direction of connecting them with cosmological scenarios in fundamental theories such as supergravity and string theory where de Sitter solutions are hard to find. One of the recent approaches in order to overcome such difficulty is the holographic renormalization group (RG) flow which is written in terms of first-order equations for a given superpotential \cite{McFadden:2009fg,Baumann:2019ghk,Kiritsis:2019wyk,Kiritsis:2013gia}. The first-order formalism has also been applied, e.g., in $f(R)$ and $f(R,T)$ gravity \cite{Afonso:2007zz,Moraes:2016gpe}, but to the best of our knowledge this is the first time where it is investigated in Horndeski cosmology. In our study, the scalar potential cannot be arbitrary since it depends on the superpotential that obeys a differential equation which constraint the possible cosmological scenarios}. Despite this, as we shall see below, there exists some restricted values of parameters that allow to describe the current acceleration of the Universe. Thus, we shall assume a dark energy dominance described by the scalar field dynamics at small redshifts. 
Let us first focus on the equation of state.

Then, by using the energy-momentum tensor of equation (\ref{11}) defined as
\begin{eqnarray}
T_{\mu\nu}=\frac{\alpha}{2k}T^{(1)}_{\mu\nu}-g_{\mu\nu}V(\phi)+\frac{\eta}{2k}T^{(2)}_{\mu\nu},
\end{eqnarray}
the $tt$ and $xx$-$yy$-$zz$-components define the effective dark energy sector with energy density and pressure 
\begin{eqnarray}
\rho_{DE}&\equiv&T_{tt}=\frac{\alpha\dot{\phi}^{2}}{4k}+\frac{V(\phi)}{2k}+\frac{9\eta H^{2}\dot{\phi}^2}{4k},\\
p_{DE}&\equiv&T_{xx}=\frac{\alpha\dot{\phi}^{2}}{4k}-\frac{V(\phi)}{2k}-\frac{\eta}{2k}\left[\frac{1}{2}\dot{\phi}^{2}(3H^{2}+2\dot{H})+2H\dot{\phi}\ddot{\phi}\right],
\end{eqnarray}
where $T_{xx}=T_{yy}=T_{zz}$. To check consistency, it is interesting to see that using the equations (\ref{1st-order-1})-(\ref{1x}), the energy density satisfies $\rho_{DE}=3H^{2}$, as expected. Thus, the dark energy equation of state is given by 
\begin{eqnarray}
\omega_{DE}=\frac{p_{DE}}{\rho_{DE}}=\frac{\alpha\dot{\phi}^{2}-2V(\phi)-2\eta\left[\frac{1}{2}\dot{\phi}^{2}(3H^{2}+2\dot{H})+2H\dot{\phi}\ddot{\phi}\right]}{\alpha\dot{\phi}^{2}+2V(\phi)+9\eta H^{2}\dot{\phi}^2}.\label{1f}
\end{eqnarray}
In terms of the dark energy density and pressure, the scalar field equation can be written in the standard form through the energy-momentum conservation, $\nabla_{\mu}T^{\mu}_{\nu}=0$, that implies
\begin{eqnarray}
\dot{\rho}_{DE}+3H(p_{DE}+\rho_{DE})=0.\label{eo}
\end{eqnarray}
This result (\ref{eo}) is in agreement with \cite{Harko:2016xip}. We may now compute another interesting cosmological quantity called decelerate parameter $q$, which indicates how the Universe expansion is accelerating and is 
given by the equation
\begin{eqnarray}
q=-\left(1+\frac{\dot{H}}{H^{2}}\right).
\end{eqnarray}
Furthermore, combining the equation (\ref{eo}) with $\rho_{DE}=3H^{2}$ we can write a useful relationship between the deceleration parameter and the equation of state as follows
\begin{eqnarray}
q=\frac{1}{2}(1+3\omega_{DE}).\label{1f1}
\end{eqnarray}
From this point we shall make use of the dimensionless redshift parameter $z$ in place of time variable $t$ in our cosmological setup since it is used to compare theoretical with observational results. The redshift parameter is defined as
\begin{eqnarray}
1+z=\frac{1}{a}.
\end{eqnarray}
Thus, time derivatives can be now expressed as
\begin{eqnarray}
\frac{d}{dt}=-H(z)(1+z)\frac{d}{dz}
\end{eqnarray}
and the first-order equations (\ref{1st-order-1}) can be simply rewritten in the new variable $z$ in the form
\begin{eqnarray}
H(z)&=&W(\phi(z)),\nonumber\\
 \phi'(z)&=&-\frac{W_{\phi}(z)}{(1+z)W(z)}. 
\end{eqnarray}
The numerical results are also computed for equation (\ref{1f}) given in terms of the superpotential
\begin{eqnarray}
\omega_{DE}=-1+\frac{2\alpha}{3}\frac{W^{2}_{\phi}}{W}+2\eta W^{2}_{\phi}+\frac{\eta W^{4}_{\phi}}{3 W^{2}}+\frac{4\eta W^{2}_{\phi}W_{\phi\phi}}{3W}.\label{1f2}
\end{eqnarray}
We summarize as follows. 
\begin{figure}[!ht]
\begin{center}
\includegraphics[scale=0.4]{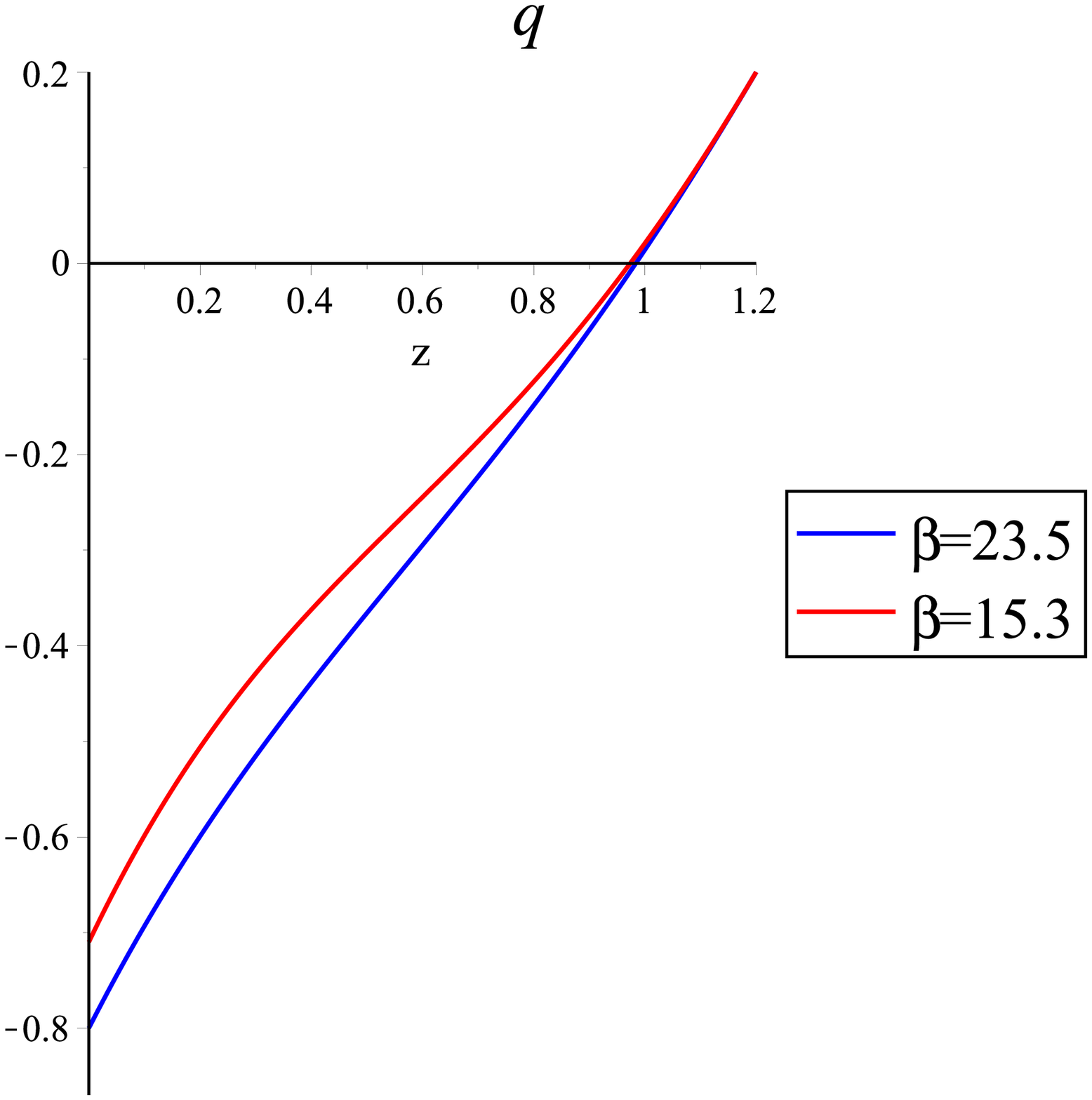}
\includegraphics[scale=0.4]{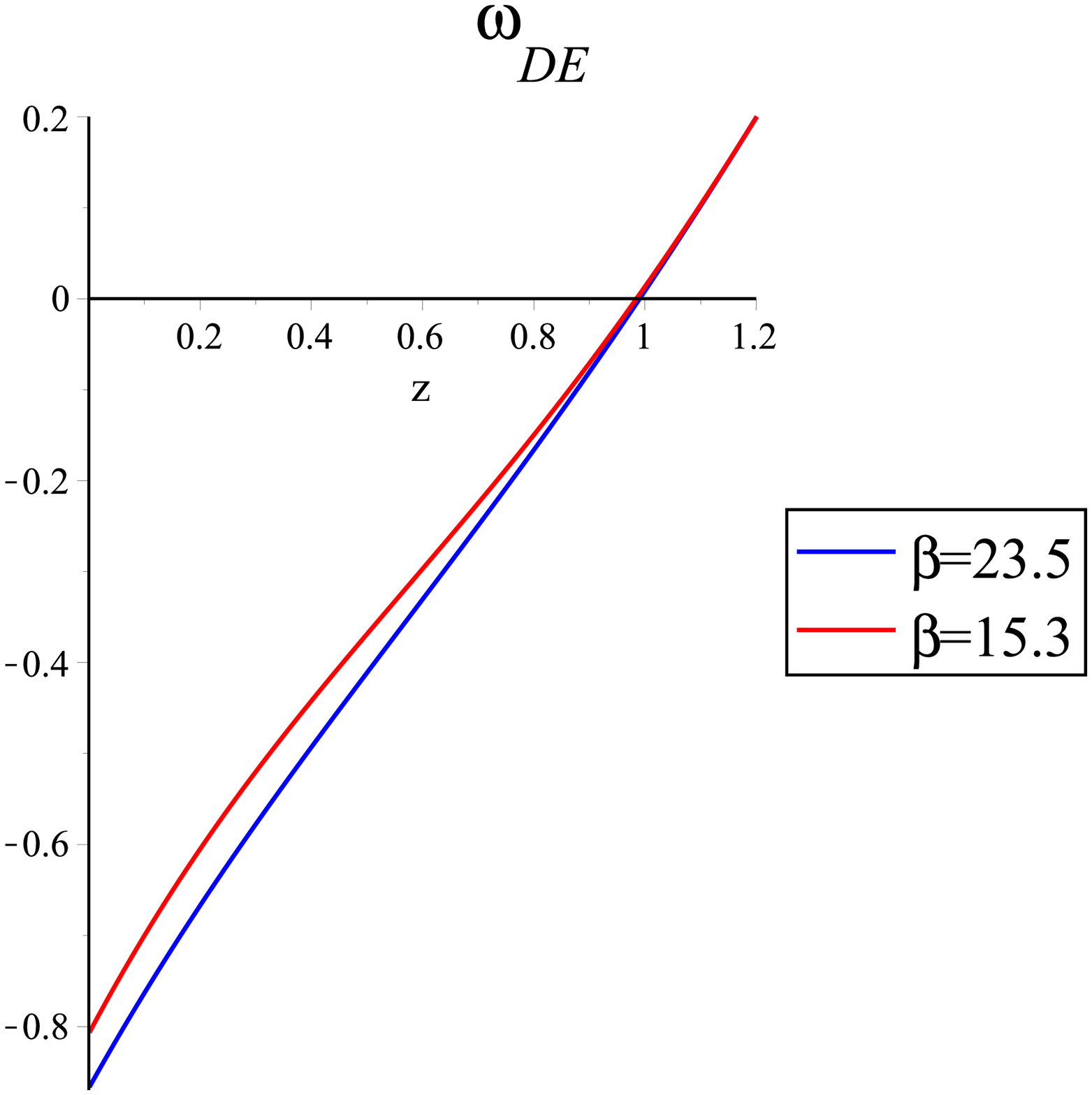}
\includegraphics[scale=0.4]{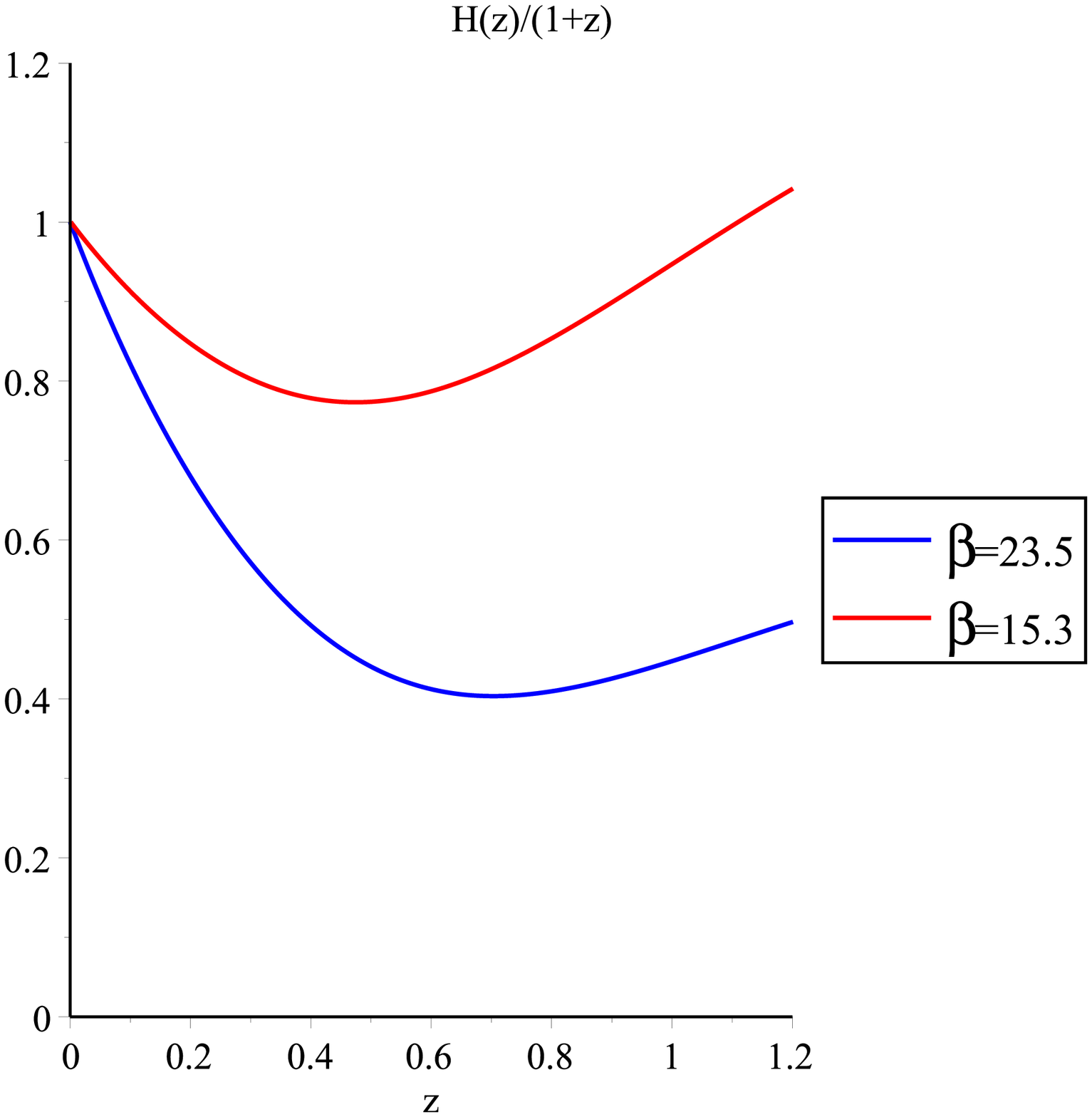}
\includegraphics[scale=0.4]{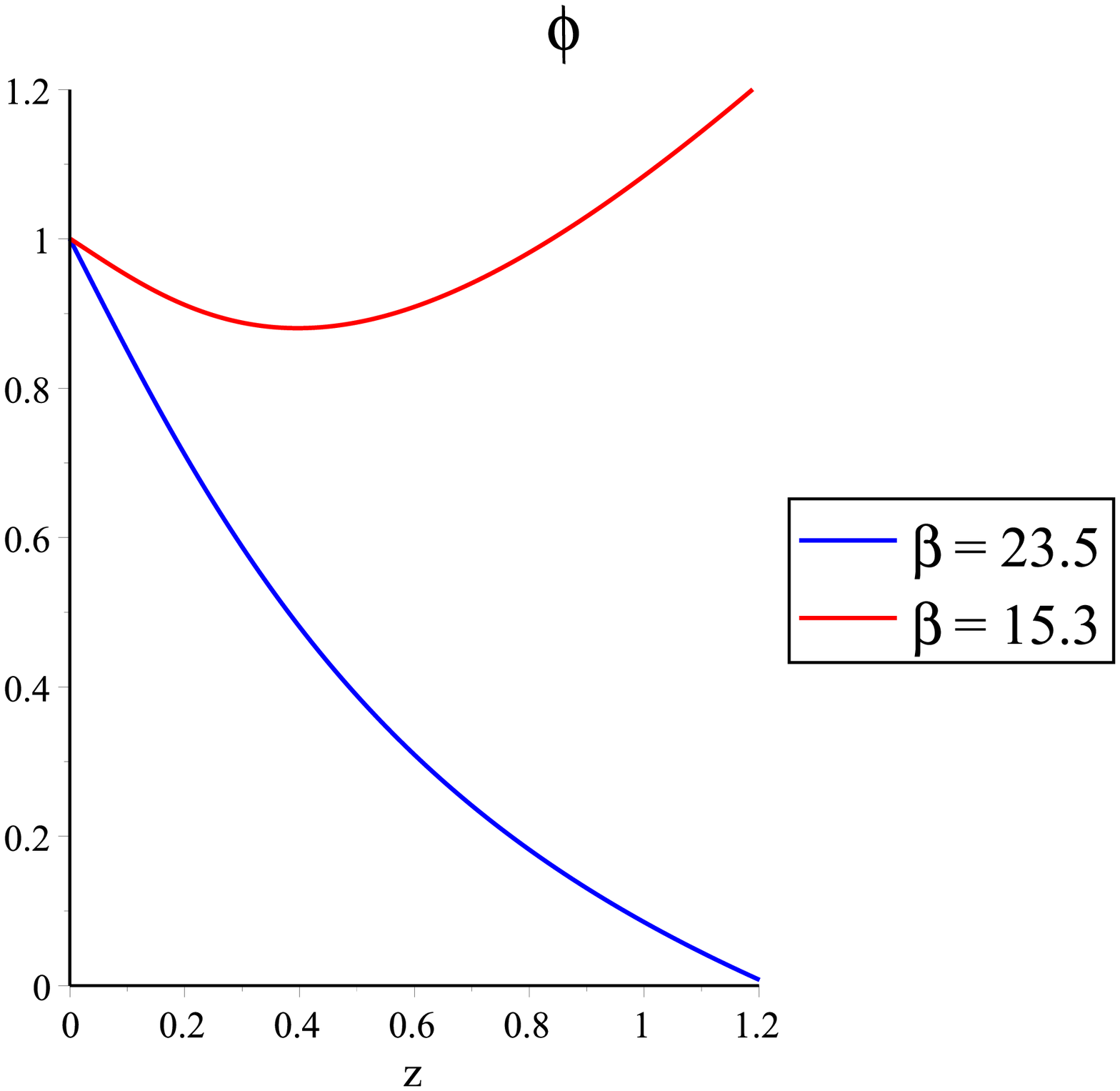}
\caption{The decelerate parameter $q$ (\ref{1f1}) (top-left) and dark energy equation of state $\omega_{DE}$ (\ref{1f2}) (top-right) for $\beta=23.5$, with $\alpha=0.06$ and $\eta=0.04$ (blue curve) and $\beta=15.3$, with $\alpha=0.08$ and $\eta=0.06$ (red curve). The the Hubble parameter $H$ (bottom-left) and scalar field $\phi$ (bottom-right) with the same values of parameters. 
All the cosmological observables are given as functions of the redshift $z$.
}
\label{h1}
\end{center}
\end{figure}
According to Fig.~\ref{h1} (top-right) the dark energy equation-of-state $\omega_{DE}$ (\ref{1f2}) acquires values near $-0.87$ (blue curve). 

{Notice that our cosmological observables here were calculated only in the scalar-tensor sector of the Horndeski gravity given by the action (\ref{10}) with no matter contribution. This sector alone is unable to take into account all phases of the Universe at high redshifts. Thus, we addressed only the dark energy (scalar) dominance at low redshifts. But investigations considering matter contribution to extension of this theory with several scalar potentials were already considered in \cite{Harko:2016xip}. Such a similar investigation in our model is out of the scope of this paper and can be addressed elsewhere. On the other hand, an inflationary phase can be found 
for a constant superpotential $W=H_0$ that satisfies (\ref{R2}). The first-order equations (\ref{1st-order-1}) give the solution $a(t)=a_0\exp{(H_0t)}$ and $\phi=const.$, which characterizes an exponential inflation with equation of state $\omega=-1$.}

Moreover from equation (\ref{1f2}) and using the exact solution of the superpotential given by equation (\ref{W-an}) we can see that the equation of state $\omega_{DE}<-1$, since in the assumed limit $\phi\approx 1$ this function is dominated by the following term
\begin{eqnarray}
\omega_{DE}=-\frac59\left[\frac{1}{\cos{(\frac{3}{2}\phi)}}\right]^{8/3}<-1
\end{eqnarray}
This is one of the advantages of the Horndeski gravity, where a phanton-like behavior is obtained even though the scalar field has canonical dynamics.  This effect is simply achieved due to the nonminimal coupling of the scalar field to gravity in the gravitational extension of the Einstein-Horndeski gravity \cite{Harko:2016xip}. 
The phantom cosmology issues are important to be addressed since recent observational data \cite{Aghanim:2018eyx} have shown the possibility of $\omega_{DE}<-1$.

The Hubble function in the Fig.~\ref{h1} (bottom-left) develops an increasing behavior between the onset at $z=0.6$ and $z=0$ (blue-curve). This fact indicates a dark energy phase dominance starting in a relatively recent time of the cosmological evolution of the Universe. The scalar field in the Fig.~\ref{h1} (bottom-right) evolves accordingly, approaching a constant at $z=0$ (blue-curve and red-curve) which represents a de Sitter  Universe \cite{Harko:2016xip}. 

{Furthermore,  as we can see from Fig.~\ref{h1} (top-left) $q(0)=-0.8$ or $-0.7$ is consistent with supernovae observations that reveals $q(0)=-0.1\pm 0.4$ \cite{Riess:1998cb}. 
}




\section{Conclusions}\label{z5}

{In the present study, we have taken the advantage of the first-order formalism in the Horndeski cosmology. This formalism applied to Horndeski gravity as well as to $f(R)$, $f(R,T)$ theories \cite{Afonso:2007zz,Moraes:2016gpe} concerns in reducing the equations of motion to first-order equations, which simplifies the solution of the problem from both analytical and numerical perspective. Besides, the first-order formalism plays an important role in RG flow in holographic cosmology \cite{McFadden:2009fg,Baumann:2019ghk,Kiritsis:2019wyk,Kiritsis:2013gia}.}
Our numerical solutions showed a good agreement with the current phase of the Universe, where the Hubble parameter as a function of the redshift has a behavior similar to the one found in \cite{Aghanim:2018eyx}.  
By using first-order formalism supported by a constrained superpotential in the Horndeski gravity for the FRW background,  we have shown by using numerical methods, that late-time cosmology is well described by the scalar field. The solutions correspond to an accelerating Universe for small redshifts, which is in agreement with the current observational data that is usually associated with a phenomenon driven by a dark energy fluid. The scalar field non-minimally coupled to the gravity sector produces kink type solutions which render a de Sitter Universe at late-time cosmology, reproducing a dark energy scenario in Horndeski gravity at first-order formalism. 


\acknowledgments

We would like to thank CNPq, CAPES, and CNPq/PRONEX, for partial financial support. FAB acknowledges support from CNPq (Grant no. 312104/2018-9). We also thank M. Rinaldi for useful discussions in the early stages of this work.


\end{document}